\pdfoutput=1

\documentclass[conference]{IEEEtran}
\usepackage[hidelinks]{hyperref}

\usepackage{cite}

\ifCLASSINFOpdf
   \usepackage[pdftex]{graphicx}	
\else
   \usepackage[dvips]{graphicx}
\fi

\usepackage{amsmath}
\usepackage{algorithmic}

\usepackage{array}

\ifCLASSOPTIONcompsoc
  \usepackage[caption=false,font=normalsize,labelfont=sf,textfont=sf]{subfig}
\else
  \usepackage[caption=false,font=footnotesize]{subfig}
\fi
\usepackage{dblfloatfix}

\usepackage{url}

\usepackage{mathabx}
\usepackage{booktabs}
\usepackage{multirow}
\usepackage{color}
\usepackage{bm}
\usepackage{xcolor}
\usepackage{pgfplots}

\graphicspath{ {images/} }
\newtheorem{definition}{Definition}[section]

\makeatletter
\newcommand\xleftrightarrow[2][]{%
  \ext@arrow 9999{\longleftrightarrowfill@}{#1}{#2}}
\newcommand\longleftrightarrowfill@{%
  \arrowfill@\leftarrow\relbar\rightarrow}
\makeatother

\begin{document}
%
\title{Predicting Anchor Links between Heterogeneous Social Networks}

\author{
\IEEEauthorblockN{Sina Sajadmanesh}
\IEEEauthorblockA{Department of Computer Eng.\\
Sharif University of Technology\\
Email: sajadmanesh@ce.sharif.edu}
\and
\IEEEauthorblockN{Hamid R. Rabiee}
\IEEEauthorblockA{Department of Computer Eng.\\
Sharif University of Technology\\
Email: rabiee@sharif.edu}
\and
\IEEEauthorblockN{Ali Khodadadi}
\IEEEauthorblockA{Department of Computer Eng.\\
Sharif University of Technology\\
Email: khodadadi@ce.sharif.edu}
}


%


\maketitle

\begin{abstract}
People usually get involved in multiple social networks to enjoy new services or to fulfill their needs. 
Many new social networks try to attract users of other existing networks to increase the number of their users. Once a user (called source user) of a social network (called source network) joins a new social network (called target network), a new inter-network link (called anchor link) is formed between the source and target networks. In this paper, we concentrated on predicting the formation of such anchor links between heterogeneous social networks. Unlike conventional link prediction problems in which the formation of a link between two existing users within a single network is predicted, in anchor link prediction, the target user is missing and will be added to the target network once the anchor link is created. To solve this problem, we use meta-paths as a powerful tool for utilizing heterogeneous information in both the source and target networks. To this end, we propose an effective general meta-path-based approach called \textit{Connector and Recursive Meta-Paths} (CRMP). By using those two different categories of meta-paths, we model different aspects of social factors that may affect a source user to join the target network, resulting in the formation of a new anchor link. Extensive experiments on real-world heterogeneous social networks demonstrate the effectiveness of the proposed method against the recent methods.

\end{abstract}


%
\IEEEpeerreviewmaketitle

\section{Introduction}
In recent years, online social networks such as Facebook, Twitter and Instagram have tremendously altered the way we communicate together. Nowadays, people use multiple social networks, simultaneously. For example, people write their daily news in Twitter while sharing their photos in Instagram. To fulfill the need for new services, modern social networks with different services and characteristics have emerged and try to attract users of other social networks. An important task for these networks is to find the users of other networks who are likely to join them, for targeted advertisements. On the other hand, the major existing social networks try to preserve their users. Therefore, it makes sense for them to find the potential outgoing users and keep them active by providing special offers. As a result, it is important to predict which users may switch from an elder social network to a newer one, in the near future.

\begin{figure*}[!t]
\centering
\subfloat[Current snapshot]{\includegraphics[width=2.5in]{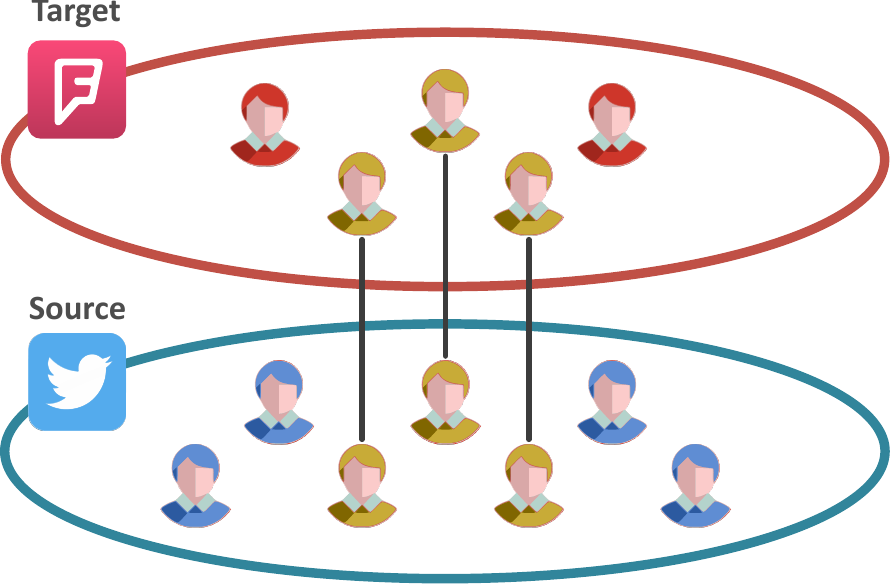}}
\hfil
\subfloat[Future snapshot]{\includegraphics[width=2.5in]{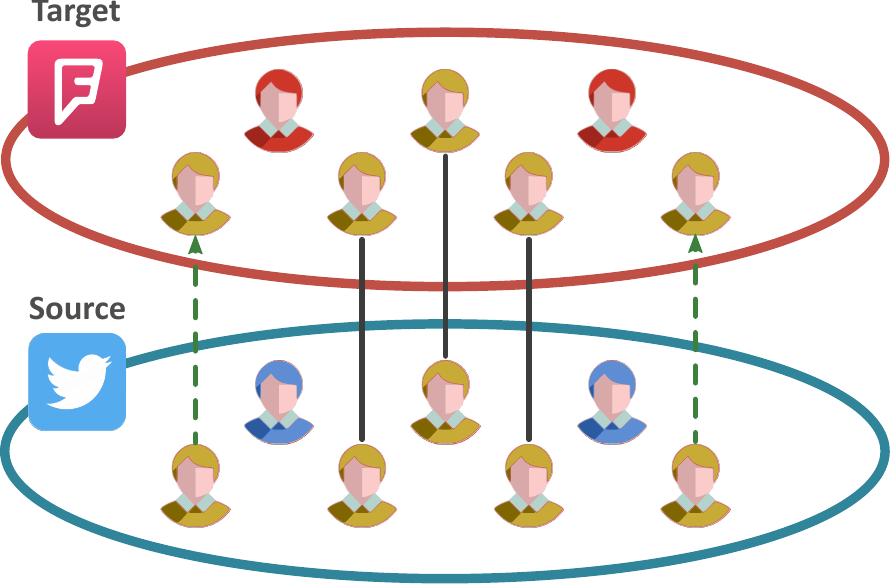}}
\caption{The formation of anchor links between source and target social networks. Blue and Red users are non-anchor users of the source and target networks, respectively. Users shown in yellow are anchor users and the links between them are anchor links. Intra-network links between users are omitted for better illustration. The figure on the left shows the current snapshot of both networks with three anchor users. The right figure shows the future snapshot in which another two non-anchor users from the source network have joined the target network and therefore two new anchor links (shown in dashed green arrows) have been formed.}
\label{fig:intro}
\end{figure*}

In general, some users may be shared between an elder social network, which we call \textit{source network}, and a newer social network, which we call \textit{target network}. These common users are known as "anchor users" and the connection between the accounts of an anchor user in different networks is abstractly referred as inter-network link called "anchor link" in the literature \cite{kong2013}. We want to find non-anchor users, i.e. unshared users, from the source network who most likely will join the target network in the future. By joining a non-anchor user from the source network to the target network causes an anchor link to be formed. This problem is equivalent to predicting anchor links which will be formed in the future. \figurename~\ref{fig:intro} illustrates the formation of anchor links.

Predicting the formation of links in a social network have been studied extensively in recent years. However, the anchor link prediction problem studied in this paper is different from traditional link prediction tasks. The main difference comes from the fact that in conventional link prediction problems, both endpoints of the missing links are presented in the same network, while in anchor link prediction problem, one of the two endpoints (which is the target one) is missing and will be added to the target network once the anchor link is created. Moreover, the problem of "anchor link inference", which has proposed in \cite{kong2013}, is another related work that is totally different from anchor link prediction. In anchor link inference problem, the goal is to discover the correspondence between accounts of the same users across multiple networks. In other words, the goal of anchor link inference is to detect anchor links which exists in reality between networks, but are unknown. On the contrary, anchor link prediction problem aims to predict formation of anchor links in the future.

Preliminary works on link prediction had only concentrated on the network structure to predict the future links, while there are many other types of information, such as locations and posts contents, that can be leveraged. In order to use the abundant information in social networks, the recent trends in link prediction have changed their focus from homogeneous networks to heterogeneous ones \cite{davis2011, sun2011co, cao2014collective, zhang2013predicting}. Unlike homogeneous networks which are only composed of single type of nodes and links, in heterogeneous networks, multiple types of nodes are related using multiple types of links. For example, alongside with users and locations that can be considered as different node types of a heterogeneous social network, social links among users, and location links between users and locations, are instances of different kinds of links. In this paper, we have focused on prediction of anchor links by leveraging heterogeneous information, and formulated both the source and target networks as two heterogeneous social networks.

The anchor link prediction problem in heterogeneous networks is a challenging issue, and some of these challenges are as follows:

\begin{itemize}
\item \textit{Lack of features}:
The extraction of heterogeneous features that are capable of utilizing heterogeneous information of both source and target networks for the anchor link prediction problem is a challenging prerequisite which has not been addressed yet.

\item \textit{Missing target node}:
The conventional link prediction approaches cannot be extended to solve the anchor link prediction problem, since the target node is missing in the later one. This issue puts forward a big challenge to be addressed.

\item \textit{Network difference problem}:
The source and target networks may have different structure and characteristics. This problem which is referred to as heterogeneity between source and target networks in \cite{wu2014learning}, is another challenge of anchor link prediction.
\end{itemize}

In this paper, we propose a meta-path-based approach called CRMP (Connector and Recursive Meta-Paths) to solve the anchor link prediction problem. We use meta-paths as a powerful tool to model the effective social factors that affect a source user to join the target network. Using network schema of both the source and target networks, and existing anchor links, we introduce two meta-path categories: connector meta-paths and recursive meta-paths. Each of these two class of meta-paths model different aspects of social factors between anchor and non-anchor users and provide a feature building mechanism that are independent of the underlying networks. We formulate the problem as a binary classification task and utilize connector and recursive meta-paths to extract a feature vector for each non-anchor user to predict formation of anchor links. Extensive experiments on real-world social networks show that the proposed CRMP method outperforms the recent relevant methods.

The rest of this paper is organized as follows. Section \ref{sec:related} provides the related works. In section \ref{sec:problem}, we present the problem formulation. We will explain the proposed method in section \ref{sec:method}. Experiment results are discussed in section \ref{sec:experiments}. Finally, we conclude the paper in section \ref{sec:conclusion}.

\section{Related Works}\label{sec:related}

\begin{table*}[!t]
\renewcommand{\arraystretch}{1.3}
\caption{Summary of Related Works}\label{table:literature}
\centering
\begin{tabular}{|l||c|c|c|c|}
\hline
Property & Link Prediction \cite{liben2007link, lu2011link, al2011survey, wang2014review} & Anchor Link Inference \cite{kong2013, zhang2015integrated} & Anchor Link Prediction \cite{wu2014learning}\\
\hline\hline
Information Sources & Single/Multiple Networks & Multiple Networks & Multiple Networks\\
Network Type & Homogeneous/Heterogeneous & Heterogeneous & Homogeneous/Heterogeneous\\
Type of Links & Intra-Network Links & Missing Anchor Links & Future Anchor Links\\
Method & Supervised/Unsupervised & Supervised/Semi-Supervised & Unsupervised\\
\hline
\end{tabular}
\end{table*}

The problem of link prediction in social networks have been studied extensively in recent years \cite{liben2007link, lu2011link, al2011survey, wang2014review}, and many distinct methods have been proposed to solve this problem. Proximity-based approaches are from the earliest methods in which a measure of similarity like common neighbors, Jaccard coefficient, and Adamic/Adar index \cite{adamic2003friends} are defined between two nodes, and missing links are being ranked based on the this measure. Another class of methods are probabilistic methods \cite{clauset2008hierarchical, airoldi2009mixed, taskar2003link} in which a probabilistic model of link creation is fitted to the network data, and prediction is performed based on the probabilities inferred from the model.
In supervised classification-based methods \cite{al2006link}, a binary classifier is trained by using the feature vectors extracted for each link in the network, to predict links.

The preliminary works have been focused on homogeneous networks containing single type of nodes and links. Recently, the use of heterogeneous information in link prediction problem has gained much attention. Sun et al. \cite{sun2011co} used a meta-path-based approach called PathPredict to predict co-authorship links in heterogeneous bibliographic networks. Yang et al. \cite{yang2012link} proposed a probabilistic method called MRIP to predict links in heterogeneous networks. Cao et al. \cite{cao2014collective} also suggested a meta-path-based method to predict multiple type of links in heterogeneous information networks. Kuo et al. \cite{kuo2013unsupervised} devised an unsupervised method using aggregative statistics for link prediction problem in heterogeneous networks.

More recent works on link prediction were focused on using multiple heterogeneous social networks to enhance performance of link prediction tasks \cite{zhang2013predicting, zhang2014transferring, zhang2014meta}. Besides, the problem of anchor link inference were suggested by Kong et al. \cite{kong2013} and a method called MNA were proposed to detect unknown anchor links across multiple heterogeneous social networks. Zhang and Yu \cite{zhang2015integrated} also proposed a method called CLF to infer both anchor and social links, simultaneously.

Meanwhile, the problem of anchor link prediction has not gained much attention in the research community, until recently. To the best of our knowledge, the closest proposed method to the anchor link prediction problem, is the CICF method of Wu et al. \cite{wu2014learning}. The main idea of CICF is to transfer knowledge only through those anchor users who behave consistently across both the source and target networks. However, they formulated the problem as a cross-domain learning task, and thus missed some other important factors such as peer influence that affects the formation of anchor links. Furthermore, they did not intend to use heterogeneous information and the task of feature extraction is delegated to the application. 

Table \ref{table:literature} shows a comparison between link prediction, anchor link inference, and anchor link prediction problems.
Besides the conventional link prediction studies, there are other works that are related to the problem of anchor link prediction. In \cite{xu2014retaining}, the most important factors that cause users to switch between social networks are investigated, empirically. They used the Push-Pull-Mooring model of migration \cite{jackson1986aspects} as a basis, and categorized different factors into push, pull, and mooring. However, they did not propose a model for prediction of user migration in social networks. In another related work \cite{backstrom2006group}, Backstrom et al. have studied the formation and evolution of communities in social networks. They have shown that for a user, both number of her friends and the associated share of her friends' activities in the target community have great impact on her to join that community. We have used the results of these studies as a background theory for the proposed method.

\section{Problem Formulation}\label{sec:problem}
In this section, we will give the formal definitions of important concepts that have been used in this paper and the formulation
of the anchor link prediction problem.

\subsection{Terminology Definition}
\begin{definition}{(Heterogeneous Social Network)}
A social network with multiple kinds of nodes and links is called a heterogeneous social network. It is represented as $G=(V,E)$ where $V=\bigcup_iV_i$ is the set of different nodes and $E=\bigcup_jE_j$ is the set of different links.
\end{definition}

\begin{definition}{(Network Schema \cite{sun2012mining})}
The schema of a heterogeneous network $G$ is a graph $S_G=(\nu, \varepsilon)$ where $\nu$ is the set of different node types and $\varepsilon$ is the set of different link types in $G$.
\end{definition}

\begin{figure}[!t]
\centering
\includegraphics[width=2.5in]{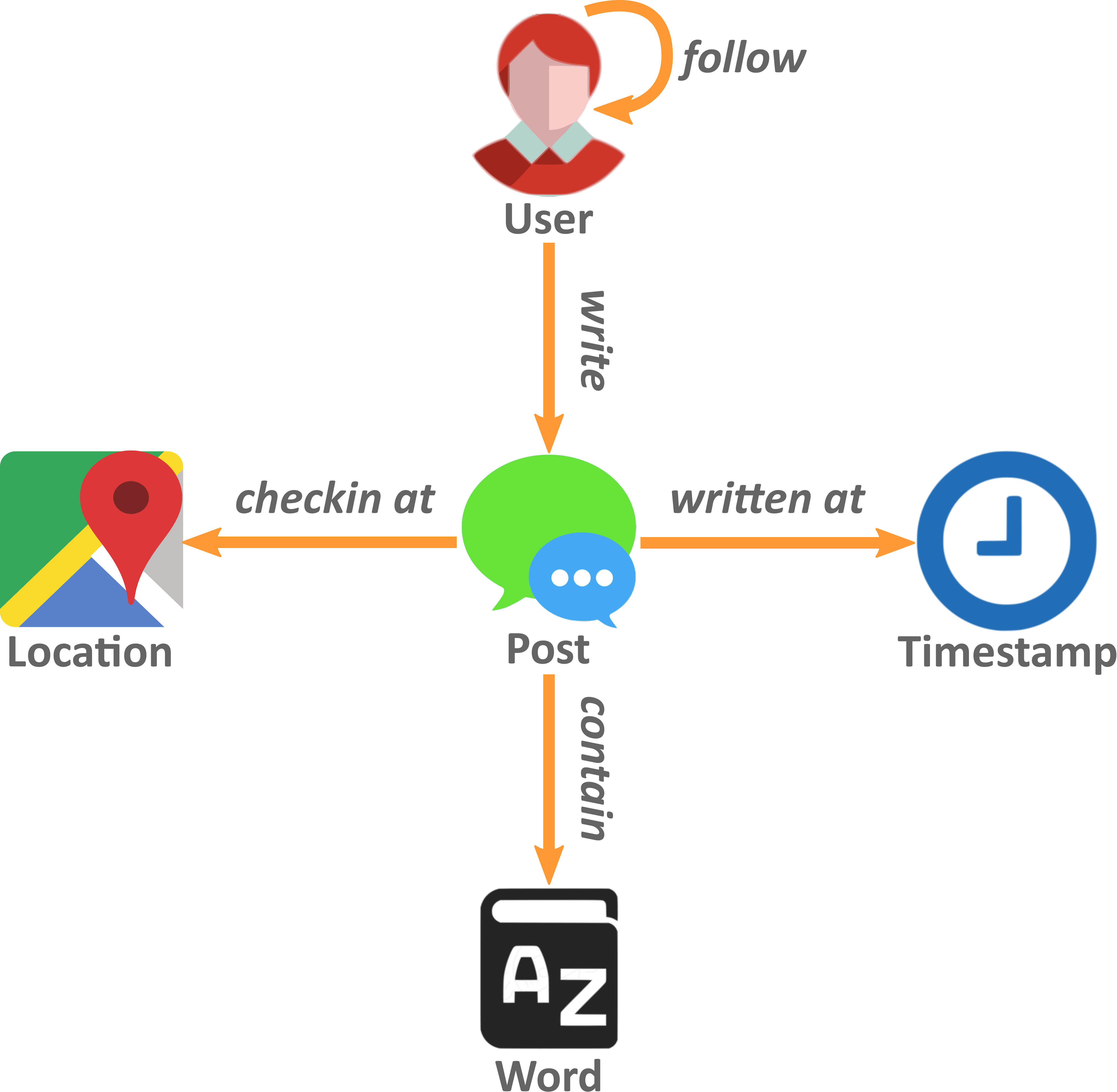}
\caption{Schema of a general heterogeneous social network}
\label{fig:schema}
\end{figure}

\figurename~\ref{fig:schema} represents the schema of a heterogeneous social network, where \textit{user}, \textit{post}, \textit{location}, \textit{word}, and \textit{time} are considered as different node types, and \textit{follow}, \textit{write}, \textit{written at}, \textit{checkin at}, and \textit{contain} are different link types. 
In this paper, we have used Twitter and Foursquare (which are both heterogeneous social networks) as the source and target networks, respectively. They have the same schema as shown in \figurename~\ref{fig:schema}. However, the proposed framework does not depend on this specific network schema and one can define his own schemas for the source and target networks depending on the social networks used in application. Using this network schema, we formulate source and target networks as $G^g=(V^g,E^g)$ where $V^g=U^g\cup P^g\cup L^g\cup T^g\cup W^g$ in which $U^g$, $P^g$, $L^g$, $T^g$, and $W^g$ denotes the set of users, posts, locations, timestamps, and words relative to network $g=s$ or $t$, respectively. Moreover, we indicate node types with capital letters (like $U$ for user type) and node instances with small letters (like $u$ for a user instance).

\begin{definition}{(Meta-Path \cite{sun2012mining})}
A meta-path is any directed path along the network schema. Formally speaking, given a network schema $S_G=(\nu, \varepsilon)$, the sequence $\nu_1\xrightarrow{\varepsilon_1}\nu_2\xrightarrow{\varepsilon_2}\dots\nu_{k-1}\xrightarrow{\varepsilon_{k-1}}\nu_k$ is a meta-path defined on $S_G$ where $\nu_i\in \nu$ and $\varepsilon_i\in \varepsilon$. A meta-path is called homogeneous if all $\nu_i$s and all $\varepsilon_i$s be the same; otherwise it is called to be a heterogeneous meta-path.
\end{definition}

\begin{definition}{(Anchor Link)}
Given the source and target networks as $G^s$ and $G^t$, and their user sets as $U^s$ and $U^t$, the link $(u_i^s, u_j^t)$ is an anchor link between $G^s$ and $G^t$ if $u_i^s\in U^s$ and $u_j^t\in U^t$ and both $u_i^s$ and $u_j^t$ belong to the same user.
\end{definition}

Despite most prior works \cite{kong2013, zhang2013predicting, zhang2014transferring, zhang2014meta, zhang2015integrated} in which anchor links are considered to be undirected, we consider them as directed links. It is important to note that the direction of anchor links only indicates the network that the user joined recently.

\begin{definition}{(Joint Network Schema)}
Given the source and target networks as $G^s$ and $G^t$, and their network schemas $S_{G^s}$ and $S_{G^t}$, we create a new schema called \textit{joint network schema} as $S_{G^{(s,t)}}=(\nu^s\cup\nu^t, \varepsilon^s\cup\varepsilon^t\cup\{anchor\})$. That means we add anchor link as a new link type to the union of the source and target network schemas.
\end{definition}

\begin{definition}{(Anchor Meta-Path)}
Given the joint network schema, we define $\alpha(U^s,U^t)=U^s\xleftrightarrow{anchor}U^t$. Unlike anchor links, we consider anchor meta-paths as undirected, which means $\alpha(U^s,U^t)=\alpha(U^t,U^s)$
\end{definition}

\subsection{Anchor Link Prediction Problem}
We now formally define the anchor link prediction problem. Given two heterogeneous social networks $G^s$ and $G^t$ as the source and target networks, besides their network schemas $S_{G^s}$ and $S_{G^t}$ respectively, the goal of anchor link prediction is to predict the formation of anchor links $(u_i^s, u_j^t)$ where the user $u_j^t$ who is called target user, is not currently in the target network and will be joined in the future.

The key challenge of anchor link prediction is the fact that the target user is missing, and the formation of anchor link with the creation of target user have to be done simultaneously. Hence, conventional link prediction approaches cannot be applied to solve the anchor link prediction problem. In the next section, we propose the CRMP method which is a supervised approach to predict anchor links directed from the source to the target network.

\section{Proposed Method}\label{sec:method}

In this section, we introduce the proposed method called \textit{Connector and Recursive Meta-Paths} to solve the anchor link prediction problem in heterogeneous social networks. We first review the required background theories that our model is built upon them. Based on these theories, we then introduce connector meta-paths and recursive meta-paths which can effectively model the theories behind the joining a source users to the target network. Using these meta-paths, a feature vector is extracted for each non-anchor user from the source network that can be used to train a binary classifier to predict formation of new anchor links.

\subsection{Background}\label{sec:method:factors}

\begin{figure}
\centering
\includegraphics[scale=1]{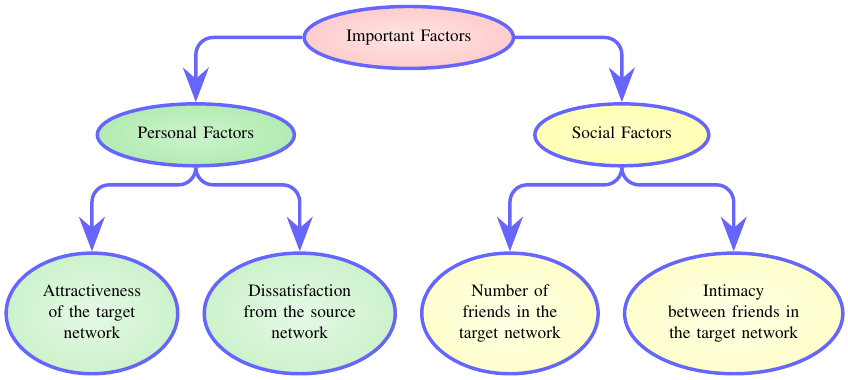}
\caption{Different factors affecting a source user to join the target network.}
\label{fig:factors}
\end{figure}

There are many factors that causes a non-anchor user from the source network to join the target network and so become an anchor one. As shown in \figurename~\ref{fig:factors}, we categorize these factors in two classes: personal and social factors.

Personal factors are those factors that are related to the user profile itself. There are two main personal factors affecting a source user to join the target network. First, the dissatisfaction with the source network; and second, the attractiveness of the target network \cite{xu2014retaining}.
On the other hand, social factors which are called \textit{peer influence} in \cite{xu2014retaining}, are the influence of friends who have already joined the target network. We consider the effect of peer influence in two folds: first, the number of user's friends who are already joined the target network; and second, the amount of activity, similarity, and connections between those friends in the target network, which we call \textit{intimacy between friends}. The more number of a source user's friends and intimacy between them in the target network, the more likely she is to join the target network.

In this paper, we just use social factors because of two reasons. First, according to \cite{xu2014retaining}, peer influence is more effective than other factors in user intension to move from a social network to a new one. Second, the modeling of social factors can be generalized over different social networks, whereas personal factors are network dependent. In the following, we introduce connector and recursive meta-paths by which we model different social factors that results in formation of new anchor links.

\subsection{Connector Meta-Paths}\label{sec:method:connector}

As we mentioned, an important social factor that affects a source user to become a target one is the number of her friends in the target network. For a source user $u^s$, the number of her followees who are in the target network can be found by counting the instances of the following meta-path starting with $u^s$:
\begin{equation*}\label{eq:neighborcount}
U^s\xrightarrow{follow}U^s\xleftrightarrow{anchor}U^t
\end{equation*}

We now claim that not only the direct followees of a source user are effective in the user intension to join the target network, but also the user would be influenced by other anchor users who are in a way \textit{similar} to that user. The reason behind this, is that the similar users often fall in the same community, and the users of a community usually perform similar actions and take similar decisions. For example, if a user belong to a community of educated people, and most of them have join a new social network, it means that the new social network is appropriate for educated people. As a result, being similar to anchor users is positively correlated with a non-anchor user's intension to join the target network. Therefore, we extend the "follow" relation in aforementioned meta-path with a similarity relationship. We call the action of extending friendship relation with similarity relationship to be \textit{similarity extension}. In order to model similarity between users, we define the notion of \textit{similarity meta-paths}.

\begin{definition}{(Similarity Meta-Path)}
Given a network schema $S_G$, a meta-path $\sigma(U,U) = U\xrightarrow{R_1}\dots\xrightarrow{R_k}U$ denoted as $U\rightsquigarrow U$, is called a similarity meta-path on $S_G$ if it is able to capture the similarity between endpoint users.
\end{definition}

We do not impose any restrictions on choosing similarity meta-paths. Depending on the real application, one can define her own meta-paths based on the schema of the source and target networks in use. For the network schema shown in \figurename~\ref{fig:schema}, we suggest the following meta-paths proposed in \cite{zhang2014meta}:

\begin{itemize}
\item $\sigma_1(U,U)=U\xrightarrow{follow}U$
\item $\sigma_2(U,U)=U\xrightarrow{follow^{-1}}U$
\item $\sigma_3(U,U)=U\xrightarrow{follow}U\xrightarrow{follow}U$
\item $\sigma_4(U,U)=U\xrightarrow{follow}U\xrightarrow{follow^{-1}}U$
\item $\sigma_5(U,U)=U\xrightarrow{follow^{-1}}U\xrightarrow{follow}U$
\item $\sigma_6(U,U)=U\xrightarrow{follow^{-1}}U\xrightarrow{follow^{-1}}U$
\item $\sigma_7(U,U)=U\xrightarrow{write}P\xrightarrow{checkin\ at}L\xrightarrow{checkin\ at^{-1}}P\xrightarrow{write^{-1}}U$
\item $\sigma_8(U,U)=U\xrightarrow{write}P\xrightarrow{written\ at}T\xrightarrow{written\ at^{-1}}P\xrightarrow{write^{-1}}U$
\item $\sigma_9(U,U)=U\xrightarrow{write}P\xrightarrow{contain}W\xrightarrow{contain^{-1}}P\xrightarrow{write^{-1}}U$
\end{itemize}

The meta-paths $\sigma_1$ to $\sigma_6$ are \textit{social similarity meta-paths} which only use social relations to capture similarity between two users. On the other hand, the meta-paths $\sigma_7$ to $\sigma_9$ which are called \textit{spatial}, \textit{temporal}, and \textit{textual similarity meta-paths}, can use heterogeneous informations such as locations, timestamps, and words to capture similarity of endpoint users, respectively.

With the definition of similarity meta-paths, we now define the notion of \textit{connector meta-paths}.

\begin{definition}{(Connector Meta-Path)}
A connector meta-path is defined as:
\begin{equation}
\Psi_i(U^s,U^t)=\sigma^s_i(U^s,U^s)\circ\alpha(U^s,U^t)
\end{equation}
where $\sigma\circ\alpha$ denotes the composition of relations $\sigma$ and $\alpha$.
\end{definition}

A connector meta-path \textit{connects} a source user to the target network using a similarity meta-path. It can be easily seen that the meta-path $U^s\xrightarrow{follow}U^s\xleftrightarrow{anchor}U^t$ which were used to count the number of followees of a source user $u^s$, who have joined the target network, is a special case of connector meta-path when we use $\sigma_1(U,U)$ as the similarity meta-path.

\subsection{Recursive Meta-Paths}

Another important social factor that affects a source user to become a member of the target network, as we talked about in section \ref{sec:method:factors}, is the level of intimacy between friends of that user in the target network. Thus, we need to model the intimacy between different users on the target network. To this end, we use the same similarity meta-paths introduced in section \ref{sec:method:connector} as a measure of intimacy between users of the target network. The intuition behind this, is that if two users be more similar to each other, and have done more common activities, then more instances of similarity meta-paths would exist between them. Based on this fact, for a non-anchor user from the source network, we model the intimacy between her friends in the target network by using the following meta-path:
\[
U^s\xrightarrow{follow}U^s\xleftrightarrow{anchor}U^t\rightsquigarrow U^t\xleftrightarrow{anchor}U^s\xrightarrow{follow^{-1}}U^s
\]

The instances of above meta-path which start and end with the same user $u^s$, means that $u^s$ has followed some users in the source network, who have already joined the target network and they are connected to each other via similarity meta-paths in the target network. Therefore, the existence of more instances of this meta-path means that a greater number of her friends are connected in the target network, and thus the likelihood of joining the user to the target network is higher. Similar to what we did in section \ref{sec:method:connector}, instead of using only direct followees of $u^s$, we extend this case using similarity extension as well which results in the definition of recursive meta-paths.

\begin{definition}{(Recursive Meta-Path)}
A recursive meta-path is defined as:
\begin{equation}
\Phi_{i,j,k}(U^s,U^s)=\Psi_i(U^s,U^t)\circ\sigma_j^t(U^t,U^t)\circ\Psi_k^{-1}(U^t,U^s)
\end{equation}
\end{definition}
 
We call this class of meta-paths "recursive" because their instances form cycles, which means for a source user $u^s$, the path begins with  $u^s$, goes to the target network and then comes back to the source network via anchor links, and finally returns to $u^s$ herself. Recursive meta-paths show that similar users to $u^s$ in the source network, are also similar to each other in the target network. As a result, existence of a great number of instances of this meta-paths means that the members of the communities in the source network, also form communities in the target network, which is positively correlated with the intension of $u^s$ to join the target network.

\subsection{Classification}
Based on the meta-paths we introduced in the previous sections, for each user in the source network, we can extract a feature vector that can be used in a supervised (or semi-supervised) classification method. The "path count" measure proposed in \cite{sun2012mining} can be used to extract features from meta-paths. More formally, let $f_{\Psi_i}$ be the feature value based on connector meta-path $\Psi_i$:
\begin{equation}
f_{\Psi_i}(u^s)=\sum_{u^t\in U^t}PC_{\Psi_i}(u^s,u^t)
\end{equation}
where $PC_{\Psi_i}$ denotes the path count of meta-path $\Psi_i$, which is the number of path instances between $u^s$ and $u^t$ following the meta-path $\Psi_i$. In this way, assuming there are $c$ different similarity meta-paths defined for the source network, the \textit{connector feature vector} for user $u^s$ would be of the form:
\begin{equation}
\Psi_F(u^s) = [f_{\Psi_1}(u^s), f_{\Psi_2}(u^s), \dots, f_{\Psi_c}(u^s)]^T
\end{equation}
Similarly, let $f_{\Phi_{i,j,k}}$ be the feature value based on recursive meta-path $\Phi_{i,j,k}$:
\begin{equation}
f_{\Phi_{i,j,k}}(u^s)=PC_{\Phi_{i,j,k}}(u^s,u^s)
\end{equation}
where $PC_{\Phi_{i,j,k}}$ denotes the path count of meta-path $\Phi_{i,j,k}$, then the \textit{recursive feature vector} for user $u^s$ would be as follows:
\begin{equation}
\Phi_F(u^s) = [f_{\Phi_{1,1,1}}(u^s), f_{\Phi_{1,1,2}}(u^s), \dots, f_{\Phi_{c,r,c}}(u^s)]^T
\end{equation}
assuming there are $c$ different similarity meta-paths defined for the source network and $r$ different similarity meta-paths defined for the target network.
The CRMP method utilizes both connector and recursive feature vectors in the form of $\left[\Psi_F^T(u^s),\Phi_F^T(u^s)\right]^T$ to train a classifier in order to predict formation of anchor links initiated by the source non-anchor users.

\section{Experiments}\label{sec:experiments}
We conducted extensive experiments to test the effectiveness of CRMP method by using real-world dataset. In this section, we first introduce the dataset and explain the experiment settings. Finally, we discuss about the experimental results.

\subsection{Datasets}
We used a heterogeneous dataset, composed of Twitter and Foursquare which have previously used in \cite{zhang2014meta, zhang2014transferring, zhang2013predicting, zhang2015integrated}. In addition to the structure of the two networks, we used locations, timestamps and text contents of the posts from both networks as heterogeneous information. During experiments, Twitter is used as the source network, and Foursquare (which is a newer network than Twitter) is used as the target network. The detailed statistics of these networks is presented in Table \ref{table:dataset}.

\begin{table}
\centering
\caption{Properties of Source and Target Networks}
\label{table:dataset}
\begin{tabular}{l l c c}
\toprule
& & \multicolumn{2}{c}{Network} \\
\cmidrule(l){3-4}
& Property & \textbf{Twitter} (Source) & \textbf{Foursquare} (Target)\\
\midrule 
& User & 5,223 & 3,456 \\ 
\# Node & Tweet/Tip & 8,205,030 & 48,585 \\ 
& Location & 257,253 & 38,861 \\ 
\midrule
& Follow & 164,919 & 16,890 \\ 
\# Link & Write & 8,205,030 & 48,585 \\ 
& Check-in & 516,149 & 48,585 \\ 
\bottomrule 
\end{tabular}
\end{table}

For each user of Twitter network, we used Twitter API to obtain the date she joined the network. Since the Foursquare API does not provide such capability, for each Foursquare user, we used the date of her first tip as the creation date of that user. A total of 3282 anchor users exist where about 1900 of them who have joined the target network after joining the source network, were used as ground truth to evaluate the performance of different methods including the proposed CRMP. The remaining anchor users were kept as existing anchor links between two networks.

\subsection{Experiment Settings}

\subsubsection{Comparison Methods}
In order to evaluate the performance of CRMP, we used the following methods:
\begin{itemize}
\item \textbf{CICF}: This method proposed in \cite{wu2014learning} is the only method presented so far for the problem of anchor link prediction.
\item \textbf{CMP}: This method is a variant of CRMP, which utilizes connector meta-paths only. This method is particularly used to analyze the effect of connector meta-paths in prediction of anchor links.
\item \textbf{RMP}: This method is a variant of CRMP, which utilizes recursive meta-paths only. This method is particularly used to analyze the effect of recursive meta-paths in prediction of anchor links.
\end{itemize}

\subsubsection{Performance Measures}
We evaluated the prediction performance of different methods under different setting using Accuracy and AUC measures. Because CICF can only output scores, for calculating Accuracy, we ranked the users according to their scores and selected top-k users as positively classified, and the remaining as negatively classified, where k is chosen equivalent to the number of positive samples.

\subsubsection{Experiment Setups}
Anchor users who have joined the source network before the target network are selected as positive users set, whose number is 1936, and non-anchor users formed the negative users set, whose number is 1941. The target node of positive users were removed from the target network. For CICF method, the optimal parameters were found by using the grid search, and path count of similarity meta-paths were used as feature vector for edges. For CRMP and its variants, Support Vector Machines (SVM) with linear kernel and default parameters were used as the base classifier. We used 5-fold cross-validation for performance evaluation: 4 folds are used for training set and remaining 1-fold is used as the test set.

\subsection{Experiment Results}
We first evaluated the prediction performance of all methods, using both homogeneous features (which are extracted using social similarity meta-paths only) and heterogeneous ones (which are extracted using social, spatial, temporal, and textual similarity meta-paths), to see the effect of different feature sets on different methods in anchor link prediction. Since the number of existing anchor links between two networks is an important factor that affects the performance of anchor link prediction, we then compared the impact of different number of existing anchor links on all methods. Next, regarding the fact that the application of anchor link prediction is for new target networks, we analyzed the effect of different degrees of newness for the target network on different methods. Finally, we evaluated the effectiveness of similarity extension to verify our assertion about extending "follow" relation with similarity relationship in both connector and recursive meta-paths.

\subsubsection{Effect of heterogeneous information}
In \figurename~\ref{fig:het}, we compared the performance of all methods using homogeneous and heterogeneous features to test the effectiveness of those features. The figure depicts that CRMP outperforms other methods under both Accuracy and AUC, using either homogeneous or heterogeneous features. Heterogeneous features increased the Accuracy of CMP, RMP, CRMP and CICF methods by about 2.5\%, 8.5\%, 3\% and 7\%, respectively. Under AUC, heterogeneous features result in an improvement of 1\% on CMP, 4\% on RMP, 3\% on CRMP, and 2.5\% on CICF. According to the results shown in the figure, using heterogeneous features, CRMP performs over 40\% better than CICF under Accuracy, and over 60\% under AUC. It can be induced that, RMP performs better than CMP, yet not better than CRMP. As a result, using connector and recursive meta-paths together is the most effective approach in anchor link prediction, while using connector meta-paths exclusively, is the least.

\newcommand{\hetplot}[1]{
\begin{tikzpicture}
\begin{axis}
[
ybar,
tiny,
width=1.9in,
enlarge x limits=0.4,
bar width=0.07in,
legend style={at={(0.5,1.3)},
anchor=north,legend columns=-1, draw=none},
symbolic x coords={CMP, RMP, CRMP, CICF},
xtick=data,
ymin=0.4,
ymajorgrids,
y tick label style={
/pgf/number format/.cd,
fixed,
fixed zerofill,
precision=2,
/tikz/.cd
},
]
\addplot+[
fill=orange,
draw=red,
error bars/.cd,y dir=both,y explicit,error mark=.] table[x=Method,y=Value, y error=Error] {results/het-#1-sc.dat};
\addplot+[
fill=cyan,
draw=blue,
error bars/.cd,y dir=both,y explicit,error mark=.] table[x=Method,y=Value, y error=Error] {results/het-#1-sc-sp-tp-tx.dat};
\legend{Homogeneous, Heterogeneous}
\end{axis}
\end{tikzpicture}
}

\begin{figure}[!t]
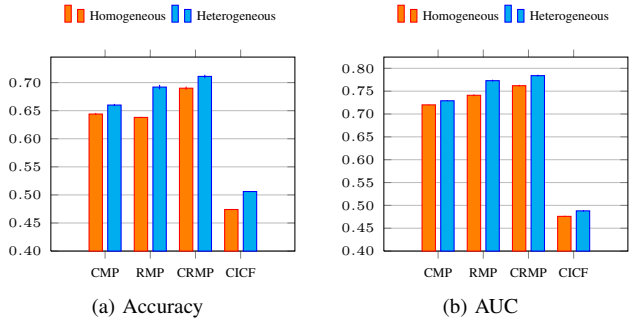

\centering
\subfloat[Accuracy]{
\hetplot{acc}
\label{fig:het:acc}}
\hfil
\subfloat[AUC]{
\hetplot{auc}
\label{fig:het:auc}}
\caption{Comparison of different methods using homogeneous and heterogeneous features with $\gamma_A=0.8$ and $\gamma_T=1$.}
\label{fig:het}
\end{figure}

\subsubsection{Effect of remaining anchor links}
We kept a fraction of randomly sampled existing anchor links under the control of parameter $\gamma_A$ which changes within $\{0.1,0.2,\dots,0.8\}$ and removed the remaining anchor links. The results are represented in \figurename~\ref{fig:ra}. As shown in the figure, it can be seen that CRMP method outperforms all other baselines consistently under both Accuracy and AUC measures. Under Accuracy and on average, CRMP is about 36.5\% better than CICF, 2.5\% better than RMP and 5.5\% better than CMP. Under AUC, CRMP is about 54\% better than CICF, 1\% better than RMP and 5\% better than CMP. We can see from the figure that as $\gamma_A$ gradually increases, the results of RMP and CRMP tend to become better, but CMP does not. This happens because as the number of anchor links increase, RMP and CRMP can leverage more information from the target network, but CMP, which does not utilize any information from the target network, remains approximately constant.

\newcommand{\anchorplot}[2]{
\begin{tikzpicture}
\begin{axis}
[
tiny,
width=1.9in,
legend style={at={(0.5,1.31)},
/tikz/every even column/.append style={column sep=0.2cm},
      anchor=north,legend columns=2, draw=none},
legend cell align=center,
ymin=0.45,ymax=#2,
xmin=0.0,xmax=0.9,
xtick={0.1, 0.2, 0.3, 0.4, 0.5, 0.6, 0.7, 0.8},
xmajorgrids,
ymajorgrids,
y tick label style={
    /pgf/number format/.cd,
        fixed,
        fixed zerofill,
        precision=2,
    /tikz/.cd
},
legend entries={CICF, CRMP, RMP, CMP},
]
\addplot[color=darkgray,thick,error bars/.cd,y dir=both,y explicit,] table[y error=e] {results/anchor-#1-cicf-ht.dat};
\addplot[color=cyan,thick,error bars/.cd,y dir=both,y explicit,] table[y error=e] {results/anchor-#1-crmp-ht.dat};
\addplot[color=orange,thick,error bars/.cd,y dir=both,y  explicit,] table[y error=e] {results/anchor-#1-rmp-ht.dat};
\addplot[color=purple,thick,error bars/.cd,y dir=both,y  explicit,] table[y error=e] {results/anchor-#1-cmp-ht.dat};
\end{axis}
\end{tikzpicture}
}

\begin{figure}[!t]
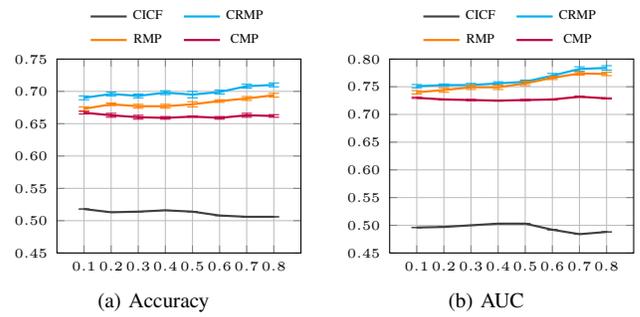

\centering
\subfloat[Accuracy]{
\anchorplot{acc}{0.75}
\label{fig:ra:acc}
}
\subfloat[AUC]{
\anchorplot{auc}{0.8}
\label{fig:ra:auc}}
\caption{Effect of remaining anchor link ratio $\gamma_A$ on performance of different methods. $\gamma_T$ is set to 1.}
\label{fig:ra}
\end{figure}

\subsubsection{Effect of newness of target network}
To represent different degrees of newness for the target network, we randomly sampled a fraction of information, including social links and tips, from the target network as available information under the control of parameter $\gamma_T$, and removed remaining information. We set $\gamma_A$ to 0.8 and changed $\gamma_T$ within $\{0.1,0.2,\dots,0.8\}$. The results are shown in \figurename~\ref{fig:rf} under the evaluation of Accuracy and AUC. Results of CMP is omitted because CMP does not utilize any information from the target network. The results show that CRMP method performs consistently better than all other methods under different degrees of newness of the target network. On average, the Accuracy achieved by CRMP is about 36\% better than CICF and 2.5\% better than RMP. Under AUC, CRMP performs about 56.5\% better than CICF and 1.5\% better than RMP. We can also see that the performance of RMP and CRMP slightly goes down while $\gamma_T$ decreases. This result shows that when the target network is new and sparse, the CRMP method can still achieve good performance.

\newcommand{\targetplot}[2]{
\begin{tikzpicture}
\begin{axis}
[
tiny,
width=1.9in,
legend style={at={(0.5,1.31)},
/tikz/every even column/.append style={column sep=0.2cm},
      anchor=north,legend columns=2, draw=none},
legend cell align=center,
ymin=0.45,ymax=#2,
xmin=0.0,xmax=0.9,
xtick={0.1, 0.2, 0.3, 0.4, 0.5, 0.6, 0.7, 0.8},
xmajorgrids,
ymajorgrids,
y tick label style={
    /pgf/number format/.cd,
        fixed,
        fixed zerofill,
        precision=2,
    /tikz/.cd
},
legend entries={CICF, CRMP, RMP},
]
\addplot[color=darkgray,thick,error bars/.cd,y dir=both,y explicit,] table[y error=e] {results/target-#1-cicf-ht.dat};
\addplot[color=cyan,thick,error bars/.cd,y dir=both,y explicit,] table[y error=e] {results/target-#1-crmp-ht.dat};
\addplot[color=orange,thick,error bars/.cd,y dir=both,y explicit,] table[y error=e] {results/target-#1-rmp-ht.dat};
\end{axis}
\end{tikzpicture}
}

\begin{figure}[!t]
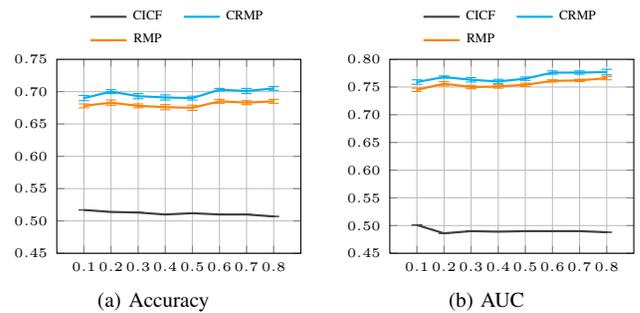

\centering
\subfloat[Accuracy]{
\targetplot{acc}{0.75}
\label{fig:rf:acc}}
\hfil
\subfloat[AUC]{
\targetplot{auc}{0.8}
\label{fig:rf:auc}}
\caption{Effect of remaining information ratio $\gamma_T$ of the target network on performance of different methods. $\gamma_A$ is set to 0.8.}
\label{fig:rf}
\end{figure}

\newcommand{\hpiplot}[1]{
\begin{tikzpicture}
\begin{axis}
[
ybar,
tiny,
width=1.9in,
enlarge x limits=0.4,
bar width=0.07in,
legend style={at={(0.5,1.3)},
anchor=north,legend columns=-1, draw=none},
symbolic x coords={CMP, RMP, CRMP},
xtick=data,
ymin=0.5,ymax=0.8,
ymajorgrids,
y tick label style={
/pgf/number format/.cd,
fixed,
fixed zerofill,
precision=2,
/tikz/.cd
},
]
\addplot+[
error bars/.cd,
y dir=both,
y explicit,error mark=.] table[x=Method,y=Value, y error=Error] {results/hpi-np-#1.dat};
\addplot+[
error bars/.cd,
y dir=both,
y explicit,error mark=.] table[x=Method,y=Value, y error=Error] {results/hpi-wp-#1.dat};
\legend{No-SE, With-SE}
\end{axis}
\end{tikzpicture}
}

\begin{figure}[!t]
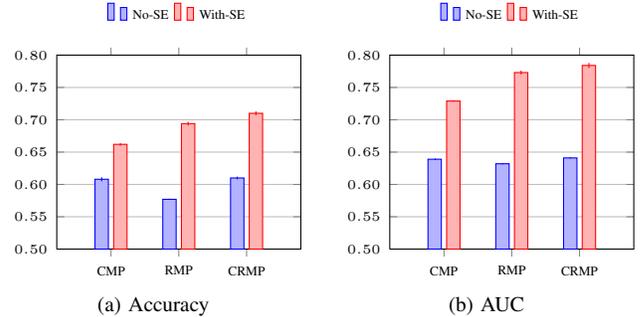

\centering
\subfloat[Accuracy]{
\hpiplot{acc}
\label{fig:hpi:acc}}
\hfil
\subfloat[AUC]{
\hpiplot{auc}
\label{fig:hpi:auc}}
\caption{Effect of similarity extension on performance of different methods with $\gamma_A=0.8$ and $\gamma_T=1$.}
\label{fig:hpi}
\end{figure}

\subsubsection{Effect of similarity extension}
We evaluated the correctness of similarity extension (SE) on CMP, RMP, and CRMP methods. By using only $\sigma_1^s$ in building connector and recursive meta-paths, we can ignore the effect of similarity extension on these methods. \figurename~\ref{fig:hpi} shows the Accuracy and AUC of No-SE and With-SE versions of CRMP and its variants. The results clearly demonstrate that applying similarity extension improved the performance of all methods considerably. Under Accuracy, similarity extension results in about 9\% improvement in CMP, 20\% improvement in RMP, and 16\% improvement in CRMP. Under AUC, similarity extension increased the result of CMP by about 14\%, RMP by about 22\% and CRMP by about 22\%. 

\section{Conclusion}\label{sec:conclusion}
In this paper, we studied the problem of anchor link prediction between heterogeneous social networks, and proposed an effective method called CRMP to solve the problem. To model the social factors affecting source users to become target ones, two class of meta-paths, \textit{connector} and \textit{recursive} meta-paths, were proposed which are built from smaller \textit{similarity} meta-paths that capture the similarity between users. By using these meta-paths, a feature vector was extracted for each non-anchor user and used in a supervised classification to predict the formation of anchor links. Extensive experiments performed on real-world dataset of Twitter and Foursquare demonstrate that CRMP outperforms the recent relevant methods.

There are many possible future directions for this work. Performing dimensionality reduction to reduce the number of features, beside using semi-supervised methods such as PU link prediction \cite{zhang2014meta}, can result in lower complexity and better prediction performance. Even though waiving personal factors brought simplicity to our model, it can be considered as a shortcoming for our approach. Thus, a potential study is to model personal factors and any other reasonable factors. Another interesting issue is to model the anchor link prediction using a temporal model to predict not only the formation of anchor links, but also the time of the formation of these links.

\section*{Acknowledgment}
We would like to thank Jiawei Zhang and Hao-Heng Chien for providing the Twitter and Foursquare dataset, and the source code of CICF, respectively. We also appreciate Reza Hadi Mogavi and Sina Jafarzadeh for their valuable comments, and ICT Innovation Center of Sharif University of Technology for its financial support.

\IEEEtriggeratref{5}


\bibliographystyle{IEEEtran}
\bibliography{IEEEabrv,references}

\begin{thebibliography}{10}
\providecommand{\url}[1]{#1}
\csname url@samestyle\endcsname
\providecommand{\newblock}{\relax}
\providecommand{\bibinfo}[2]{#2}
\providecommand{\BIBentrySTDinterwordspacing}{\spaceskip=0pt\relax}
\providecommand{\BIBentryALTinterwordstretchfactor}{4}
\providecommand{\BIBentryALTinterwordspacing}{\spaceskip=\fontdimen2\font plus
\BIBentryALTinterwordstretchfactor\fontdimen3\font minus
  \fontdimen4\font\relax}
\providecommand{\BIBforeignlanguage}[2]{{%
\expandafter\ifx\csname l@#1\endcsname\relax
\typeout{** WARNING: IEEEtran.bst: No hyphenation pattern has been}%
\typeout{** loaded for the language `#1'. Using the pattern for}%
\typeout{** the default language instead.}%
\else
\language=\csname l@#1\endcsname
\fi
#2}}
\providecommand{\BIBdecl}{\relax}
\BIBdecl

\bibitem{kong2013}
X.~Kong, J.~Zhang, and P.~S. Yu, ``Inferring anchor links across multiple
  heterogeneous social networks,'' in \emph{Proceedings of the 22Nd ACM
  International Conference on Information \& Knowledge Management}, ser. CIKM
  '13.\hskip 1em plus 0.5em minus 0.4em\relax New York, NY, USA: ACM, 2013, pp.
  179--188.

\bibitem{davis2011}
D.~Davis, R.~Lichtenwalter, and N.~V. Chawla, ``Multi-relational link
  prediction in heterogeneous information networks,'' in \emph{Advances in
  Social Networks Analysis and Mining (ASONAM), 2011 International Conference
  on}, July 2011, pp. 281--288.

\bibitem{sun2011co}
Y.~Sun, R.~Barber, M.~Gupta, C.~C. Aggarwal, and J.~Han, ``Co-author
  relationship prediction in heterogeneous bibliographic networks,'' in
  \emph{Proceedings of the 2011 International Conference on Advances in Social
  Networks Analysis and Mining}, ser. ASONAM '11.\hskip 1em plus 0.5em minus
  0.4em\relax Washington, DC, USA: IEEE Computer Society, 2011, pp. 121--128.

\bibitem{cao2014collective}
B.~Cao, X.~Kong, and P.~S. Yu, ``Collective prediction of multiple types of
  links in heterogeneous information networks,'' in \emph{Proceedings of the
  2014 IEEE International Conference on Data Mining}, ser. ICDM '14.\hskip 1em
  plus 0.5em minus 0.4em\relax Washington, DC, USA: IEEE Computer Society,
  2014, pp. 50--59.

\bibitem{zhang2013predicting}
J.~Zhang, X.~Kong, and P.~S. Yu, ``Predicting social links for new users across
  aligned heterogeneous social networks,'' in \emph{Data Mining (ICDM), 2013
  IEEE 13th International Conference on}, Dec 2013, pp. 1289--1294.

\bibitem{wu2014learning}
S.-H. Wu, H.-H. Chien, K.-h. Lin, and P.~Yu, ``Learning the consistent behavior
  of common users for target node prediction across social networks,'' in
  \emph{Proceedings of the 31st International Conference on Machine Learning
  (ICML-14)}, 2014, pp. 298--306.

\bibitem{liben2007link}
D.~Liben-Nowell and J.~Kleinberg, ``The link prediction problem for social
  networks,'' in \emph{Proceedings of the Twelfth International Conference on
  Information and Knowledge Management}, ser. CIKM '03.\hskip 1em plus 0.5em
  minus 0.4em\relax New York, NY, USA: ACM, 2003, pp. 556--559.

\bibitem{lu2011link}
L.~Lü and T.~Zhou, ``Link prediction in complex networks: A survey,''
  \emph{Physica A: Statistical Mechanics and its Applications}, vol. 390,
  no.~6, pp. 1150 -- 1170, 2011.

\bibitem{al2011survey}
M.~A. Hasan and M.~J. Zaki, \emph{Social Network Data Analytics}.\hskip 1em
  plus 0.5em minus 0.4em\relax Boston, MA: Springer US, 2011, ch. A Survey of
  Link Prediction in Social Networks, pp. 243--275.

\bibitem{wang2014review}
T.~Wang and G.~Liao, ``A review of link prediction in social networks,'' in
  \emph{Management of e-Commerce and e-Government (ICMeCG), 2014 International
  Conference on}, Oct 2014, pp. 147--150.

\bibitem{zhang2015integrated}
J.~Zhang and P.~S. Yu, ``Integrated anchor and social link predictions across
  social networks,'' in \emph{Proceedings of the 24th International Conference
  on Artificial Intelligence}, ser. IJCAI'15.\hskip 1em plus 0.5em minus
  0.4em\relax AAAI Press, 2015, pp. 2125--2131.

\bibitem{adamic2003friends}
L.~A. Adamic and E.~Adar, ``Friends and neighbors on the web,'' \emph{Social
  Networks}, vol.~25, no.~3, pp. 211 -- 230, 2003.

\bibitem{clauset2008hierarchical}
A.~Clauset, C.~Moore, and M.~Newman, ``Hierarchical structure and the
  prediction of missing links in networks,'' \emph{Nature}, vol. 453, no. 7191,
  pp. 98--101, 2008, cited By 642.

\bibitem{airoldi2009mixed}
E.~M. Airoldi, D.~M. Blei, S.~E. Fienberg, and E.~P. Xing, ``Mixed membership
  stochastic blockmodels,'' in \emph{Advances in Neural Information Processing
  Systems 21}, D.~Koller, D.~Schuurmans, Y.~Bengio, and L.~Bottou, Eds.\hskip
  1em plus 0.5em minus 0.4em\relax Curran Associates, Inc., 2009, pp. 33--40.

\bibitem{taskar2003link}
B.~Taskar, M.~fai Wong, P.~Abbeel, and D.~Koller, ``Link prediction in
  relational data,'' in \emph{Advances in Neural Information Processing Systems
  16}, S.~Thrun, L.~K. Saul, and B.~Sch\"{o}lkopf, Eds.\hskip 1em plus 0.5em
  minus 0.4em\relax MIT Press, 2004, pp. 659--666.

\bibitem{al2006link}
M.~A. Hasan, V.~Chaoji, S.~Salem, and M.~Zaki, ``Link prediction using
  supervised learning,'' in \emph{In Proc. of SDM 06 workshop on Link Analysis,
  Counterterrorism and Security}, 2006.

\bibitem{yang2012link}
Y.~Yang, N.~Chawla, Y.~Sun, and J.~Hani, ``Predicting links in multi-relational
  and heterogeneous networks,'' in \emph{Data Mining (ICDM), 2012 IEEE 12th
  International Conference on}, Dec 2012, pp. 755--764.

\bibitem{kuo2013unsupervised}
T.-T. Kuo, R.~Yan, Y.-Y. Huang, P.-H. Kung, and S.-D. Lin, ``Unsupervised link
  prediction using aggregative statistics on heterogeneous social networks,''
  in \emph{Proceedings of the 19th ACM SIGKDD International Conference on
  Knowledge Discovery and Data Mining}, ser. KDD '13.\hskip 1em plus 0.5em
  minus 0.4em\relax New York, NY, USA: ACM, 2013, pp. 775--783.

\bibitem{zhang2014transferring}
J.~Zhang, X.~Kong, and P.~S. Yu, ``Transferring heterogeneous links across
  location-based social networks,'' in \emph{Proceedings of the 7th ACM
  International Conference on Web Search and Data Mining}, ser. WSDM '14.\hskip
  1em plus 0.5em minus 0.4em\relax New York, NY, USA: ACM, 2014, pp. 303--312.

\bibitem{zhang2014meta}
J.~Zhang, P.~S. Yu, and Z.-H. Zhou, ``Meta-path based multi-network collective
  link prediction,'' in \emph{Proceedings of the 20th ACM SIGKDD International
  Conference on Knowledge Discovery and Data Mining}, ser. KDD '14.\hskip 1em
  plus 0.5em minus 0.4em\relax New York, NY, USA: ACM, 2014, pp. 1286--1295.

\bibitem{xu2014retaining}
Y.~C. Xu, Y.~Yang, Z.~Cheng, and J.~Lim, ``Retaining and attracting users in
  social networking services: An empirical investigation of cyber migration,''
  \emph{The Journal of Strategic Information Systems}, vol.~23, no.~3, pp. 239
  -- 253, 2014.

\bibitem{jackson1986aspects}
J.~Jackson, \emph{Migration}, ser. Aspects of Modern Sociology Series.\hskip
  1em plus 0.5em minus 0.4em\relax Longman, 1986.

\bibitem{backstrom2006group}
L.~Backstrom, D.~Huttenlocher, J.~Kleinberg, and X.~Lan, ``Group formation in
  large social networks: Membership, growth, and evolution,'' in
  \emph{Proceedings of the 12th ACM SIGKDD International Conference on
  Knowledge Discovery and Data Mining}, ser. KDD '06.\hskip 1em plus 0.5em
  minus 0.4em\relax New York, NY, USA: ACM, 2006, pp. 44--54.

\bibitem{sun2012mining}
Y.~Sun and J.~Han, \emph{Mining Heterogeneous Information Networks: Principles
  and Methodologies}.\hskip 1em plus 0.5em minus 0.4em\relax Morgan \& Claypool
  Publishers, 2012.

\end{thebibliography}
%



\end{document}